# A novel joint location-scale testing framework for improved detection of variants with main or interaction effects

David Soave,[1,2] Andrew D. Paterson,[1,2] Lisa J. Strug,[1,2] Lei Sun[1,3]*

[1]Division of Biostatistics, Dalla Lana School of Public Health, University of Toronto, Toronto, Ontario, M5T 3M7, Canada;

[2] Program in Genetics and Genome Biology, Research Institute, The Hospital for Sick Children, Toronto, Ontario, M5G 0A4, Canada;

[3]Department of Statistical Sciences, University of Toronto, Toronto, Ontario, M5S 3G3, Canada.

*Correspondence to:

Lei Sun
Department of Statistical Sciences
100 St George Street
University of Toronto
Toronto, ON M5S 3G3, Canada.
E-mail: sun@utstat.toronto.edu

**Abstract**

Gene-based, pathway and other multivariate association methods are now commonly implemented in whole-genome scans, but paradoxically, do not account for GxG and GxE interactions which often motivate their implementation in the first place.  Direct modeling of potential interacting variables, however, can be challenging due to heterogeneity, missing data or multiple testing.   Here we propose a novel and easy-to-implement joint location-scale (JLS) association testing procedure that can account for complex genetic architecture without explicitly modeling interaction effects, and is suitable for large-scale whole-genome scans and meta-analyses.  We focus on Fisher's method and use it to combine evidence from the standard location/mean test and the more recent scale/variance test (JLS-Fisher), and we describe its use for single-variant, gene-set and pathway association analyses. We support our findings with analytical work and large-scale simulation studies.  We recommend the use of this analytic strategy to rapidly and correctly prioritize susceptibility loci across the genome for studies of complex traits and complex secondary phenotypes of `simple' Mendelian disorders.

**Introduction**

Identifying the genetic architecture of complex traits requires analytic strategies that move beyond single-variant association tests. Multivariate analyses such as gene-based, gene-by-gene interaction (GxG), gene-by-environment interaction (GxE), gene-set and pathway analyses are now commonly implemented [Cordell 2009; Cornelis, et al. 2012; Maciejewski 2013; Mukherjee, et al. 2012; Sun, et al. 2012], yet, surprisingly, one rarely sees GxG or GxE explicitly accounted for within gene-set and pathway analyses [Ritchie, et al. 2001]. Specifying interacting variables that likely differ between genes in gene-sets and pathways is not straightforward. The interacting exposure variables (termed $E$ hereafter) could be environmental factors, SNPs or haplotypes from the same region or at other susceptibility loci. Missing or incorrect information on interacting factors as well as associated computational burden may also limit more comprehensive surveys of the whole-genome for disease association. Here we provide an easy-to-implement, straightforward solution to exploit potential GxG and GxE in gene-set and pathway analyses.

For a biallelic SNP, an underlying genetic interaction effect (either GxG or GxE) on a quantitative phenotype will lead to differences in phenotypic distributions across the three genotypes, potentially leading to differences in phenotypic variance (scale) [Pare, et al. 2010]. In light of this, Levene's scale test of equality of variance [Levene 1960] has been proposed as a method of prioritizing SNPs for subsequent GxG and GxE studies [Deng, et al. 2013; Pare, et al. 2010], in contrast to the standard mean (location) test (i.e. testing for phenotypic differences in mean across genotypes). The advantage of using variance testing to incorporate potential GxG or GxE is that exposures need not be specified nor measured; and the enormous multiple testing burden of formally examining all possible pair-wise interactions is removed. The limitation of this approach, however, is that it has no power to detect SNPs displaying main effects only.

Focusing on single-SNP analysis, Aschard, et al. [2013] recently proposed a distribution test that compares the percentiles of phenotypic values between genotypes; capturing either mean or variance differences, or both. While this approach comprehensively evaluates the phenotypic distribution between genotypes, it sacrifices statistical power when an (approximately) normally distributed trait is sufficiently summarized by its mean and variance. Furthermore, the distribution test statistic does not follow a known distribution and p-value estimation requires permutation-based methods that can be computationally challenging in a genome-wide scan setting.

More recently, Cao, et al. [2014] considered a joint test of mean or variance differences using a full likelihood approach based on normal linear regression models. The statistic from the likelihood ratio test (LRT) follows an asymptotic chi-square distribution under the null hypothesis. The LRT approach can increase power but is also known to be more sensitive to model assumptions such as normality. In response, Cao, et al. [2014] proposed a parametric bootstrap method to calculate 'honest' p-values at the cost of computational efficiency. For both the LRT and distribution methods, it is difficult to implement them for multivariate (e.g. gene-based, gene-set and pathway) analyses (see Discussion).

Here we propose a novel and easy to implement joint location-scale (JLS) testing framework that tests the null hypothesis of equal mean and equal variance between genotypes simultaneously, by combining evidence from the individual location-only and scale-only tests, focusing on Fisher's method of combining association evidence (JLS-Fisher). The proposed method detects association in the presence of underlying genetic main and/or interaction effects, *without* specifying the precise mechanism of the association; it allows any type of individual location and scale tests to be combined, making it particularly useful for gene-based, gene-set and pathway analysis.

The recent work of Dudbridge and Fletcher [2014] showed that many previously reported GxE or GxG interactions could be spurious/synthetic if there were dependence between the variables. However, their findings are also *"reassuring because they mean that when a marker-exposure interaction exists, the marker must be associated with a causal variant."* Thus, the potential synthetic interactions, though not biologically meaningful, can be leveraged and incorporated into more powerful association tests such as the proposed JLS methods that rapidly screen the whole genome to correctly prioritize susceptibility loci for further examination.

Through extensive simulation analyses, we demonstrate that the proposed JLS method has good type 1 error control with improved power compared to the individual location-only and scale-only testing options, and, in contrast to the recently proposed distribution and LRT joint testing methods [Aschard, et al. 2013; Cao, et al. 2014], can be easily implemented in the context of gene-set and pathway analyses.

**Methods**

*Genetic Model*

We considered $Y$ to be a quantitative trait of interest and $G$ to be the minor allele count for the SNP under investigation ($G$ = 0, 1 or 2); the additive assumption was not critical to the method development. For X chromosome SNPs, female and male genotypes were analyzed together and coded as $G$ = 0, 1 or 2 and $G$ = 0 or 2, respectively. In addition, we assumed the presence of an exposure variable ($E$) interacting with the genetic factor. This exposure $E$ could reflect continuous or categorical measures of environmental or genetic background (e.g. age, smoking status, genotype at another SNP or a haplotype). The underlying true genetic model may include main effects of both $G$ ($\beta_G$) and $E$ ($\beta_E$) on $Y$, as well as the interaction effect ($\beta_{GE}$):

$$\text{(True model)} \quad Y \sim \beta_0 + \beta_G G + \beta_E E + \beta_{GE} GE + \varepsilon. \quad (1)$$

Traditionally, we assume that the trait $Y$ is (approximately) normally distributed with unit variance conditional upon $G$ and $E$, in other words, Var($Y|G = g, E = e$) = 1 and $\varepsilon \sim \mathcal{N}(0, 1)$.

When considering only $G$, the working model would reduce to

$$\text{(Working model)} \quad Y \sim \beta_0 + \beta_G G + \varepsilon_G. \quad (2)$$

Pare, et al. [2010] showed that the conditional variance of $Y$ conditional on $G$ alone could be expressed as $\sigma_G^2$ = Var($Y|G = g$) = $(\beta_E + \beta_{GE} g)^2 + 1$. Thus if an interaction effect was present (i.e. $\beta_{GE} \neq 0$), the trait variance would differ between genotypes.

*Joint Location-Scale (JLS) Testing Procedures for Single-SNP and Gene-Set Analyses*

Our proposed JLS testing framework, based on the working model of equation (2), tests the following null hypothesis,

$$H_0^{joint}: \beta_G = 0 \text{ and } \sigma_i = \sigma_j \text{ for all } i \neq j, i, j = 0, 1, 2.$$

The alternative hypothesis of interest is

$$H_1^{joint}: \beta_G \neq 0 \text{ or } \sigma_i \neq \sigma_j \text{ for some } i \neq j.$$

For a SNP $G$ under study, different JLS test statistics can be considered. Let $p_L$ be the p-value for the location test of choice (i.e. testing $H_0^{location}: \beta_G = 0$ using, for example, ordinary least-squares regression), and $p_S$ be the p-value for the scale test of choice (i.e. testing $H_0^{scale}: \sigma_i = \sigma_j$ for all $i \neq j$ using, for example, Levene's test). We first consider Fisher's method (JLS-Fisher) to combine the association evidence from the individual location and scale tests. The JLS-Fisher statistic is defined as

$$W_F = -2\log(p_L) - 2\log(p_S).$$

Large values of $W_F$ correspond to small values of $p_L$ and/or $p_S$ and provide evidence against the null $H_0^{joint}$. If $p_L$ and $p_S$ are independent under $H_0^{joint}$, $W_F$ is distributed as a $\chi_4^2$ random variable. Although Fisher's method here is used to combine evidence from two tests applied to the same sample, the assumption of independence between $p_L$ and $p_S$ under $H_0^{joint}$ is shown to hold theoretically for a normally distributed trait (Lemma below), as well as empirically for approximately normally distributed traits in applications (results not shown).

One can also consider the minimum p-value (JLS-minP) approach, or various alternatives based on combining the individual test statistics themselves with or without weights [Derkach, et al. 2013; Owen 2009]. The JLS-minP statistic is defined as

$$W_M = \min(p_L, p_S).$$

If $p_L$ and $p_S$ are independent under $H_0^{joint}$, $W_M$ is distributed as a Beta random variable (with shape parameters 1 and 2) where small values of $W_M$ correspond to small values of $p_L$ or $p_S$ and evidence against the null.

The chosen JLS test statistic (e.g. $W_F$) for single-SNP analysis can then be used for implementing gene-set or pathway analysis in a direct fashion. Assume that $J$

SNPs have been annotated to a gene-set of interest. For each SNP $j$, the JLS test statistic, $W_{F,j}$, is first obtained and then the association evidence can be aggregated across the SNPs by considering, for example, the sum statistic, $\sum_j W_{F,j}$ [Sun, et al. 2012]. To account for LD between SNPs, the overall association evidence can be evaluated using a phenotype-permutation approach where the empirical p-value is the proportion of $K$ permutation replicates with sum statistics more extreme than the observed value. Because this multivariate method analyzes all $J$ SNPs simultaneously, the number of permutations need not be exceedingly large and $K =$ 10,000 provides accurate estimates for p-values in the range of 0.05. If multiple gene-sets are of interest, more replicates would be required to adjust for the corresponding multiple hypothesis testing.

*Lemma - Independence of Location-only and Scale-only Test Statistics for Normally Distributed Trait*

Let $T_{Location} = \frac{\hat{\beta}_1}{S/\sqrt{S_{xx}}}$ be our *location*-only test statistic, testing the linear effect of $x$ on $Y$ where $S^2 = \frac{1}{n-2}\sum(y_i - \hat{\beta}_0 - \hat{\beta}_1 x_i)^2$, $\hat{\beta}_0 = \bar{y} - \hat{\beta}_1 \bar{x}$, $\hat{\beta}_1 = S_{xy}/S_{xx}$, $S_{xy} = \sum(x_i - \bar{x})(y_i - \bar{y})$, and $S_{xx} = \sum(x_i - \bar{x})^2$ (in equation (2) of methods section $x = G$) and Let $T_{Scale}$ be our scale-only test statistic, here defined as Levene's test statistic for equality of variances [Levene 1960].

**Lemma** For the conditional normal model $Y_i \sim \mathcal{N}(\beta_0 + \beta_1 x_i, \sigma_{x_i}^2)$, where $x_i = 0, 1$ or 2, let $\sigma_0^2 = \sigma_1^2 = \sigma_2^2$. Then $T_{Location}$ and $T_{Scale}$ are independent.

Proof. For fixed $x$, $Y$ is normally distributed with constant variance $\sigma^2$ and mean $E[Y|x] = \beta_0 + \beta_1 x$. The joint density of $Y$ is

$$(2\pi\sigma^2)^{-n/2} \exp\left[-\frac{1}{2\sigma^2}\sum(y_i - \beta_0 - \beta_1 x_i)^2\right]$$

This is an exponential family with 3 parameters $\boldsymbol{\theta} = (\theta_1, \theta_2, \theta_3) = (\frac{\beta_1}{\sigma^2}, \frac{-1}{2\sigma^2}, \frac{\beta_0}{\sigma^2})$ for which the sufficient statistics $\boldsymbol{T} = (T_1, T_2, T_3) = (\sum x_i y_i, \sum y_i^2, \sum y_i)$ are complete. Since $T_{Scale}$ is approximately distributed as an $F_{3-1, N-3}$ variable, $T_{Scale}$ does not depend on $\boldsymbol{\theta}$ (i.e. $T_{Scale}$ is ancillary for $\boldsymbol{\theta}$). Thus, it follows that $\boldsymbol{T}$ is independent of $T_{Scale}$ (See Lehmann and Romano [2005], p.152). Since $T_{Location}$ is a function of $\boldsymbol{T}$, $T_{Location}$ and $T_{Scale}$ are independent.

Note that the proof of independence holds regardless of the version of Levene's test statistic chosen, provided that the approximation to the $F$ distribution (or some other distribution not depending on $\boldsymbol{\theta}$) is justifiable. Similar statements of independence with analogous proofs can be obtained for other choices of location test statistics such as the analysis of variance (ANOVA) $F$-statistic.

**Simulation Studies**

*Simulation Models*

We followed the exact simulation models and power analyses performed by Aschard, et al. [2013] to evaluate the performance of the proposed JLS tests and compare them with the results from the individual location-only and scale-only tests, the joint LRT [Cao, et al. 2014] and the distribution test [Aschard, et al. 2013]. The following 3 models were used to simulate the data:

Model (i)   $E[Y] = \beta_G G + \beta_{E1} E_1 + \beta_{GE1} G \cdot E_1$

Model (ii)  $E[Y] = \beta_{E1} E_1 + \beta_{E2} E_2 + \beta_{GE1} G \cdot E_1 + \beta_{GE2} G \cdot E_2$

Model (iii) $E[Y] = \beta_{GE1} G \cdot E_1$

For all three models, the observed genetic variant ($G$) was coded additively with minor allele frequency (MAF) of 0.3. $Y$ was simulated from models with varying effects ($\beta$s) and residual variation ($\varepsilon$) following a Normal distribution (mean = 0, standard deviation = 1).

Model (i) is analogous to equation (1) where $Y$ is a function of main effects of both $G$ and $E_1$, and an interaction effect between $G$ and $E_1$. The unobserved exposure variable $E_1$ was binary with frequency 0.3. The main genetic effect $\beta_G$ took on values of 0.01, 0.05 and 0.1, while the interaction effect $\beta_{GE1}$ was varied between -1 and 1 by a grid of 0.1. The main exposure effect $\beta_{E1}$ was fixed at 0.3 when $\beta_{GE1}$ was positive and -0.3 when $\beta_{GE1}$ was negative.

For model (ii), $Y$ was a function of main effects due to two unobserved exposures ($E_1$ and $E_2$; both binary with frequency 0.3) and interaction effects between the exposures and $G$. $\beta_{GE1}$ was always positive and less than 1, while $\beta_{GE2}$ was varied between -1 and 1 by a grid of 0.1. $\beta_{E1}$ was fixed at 0.3, while $\beta_{E2}$ was fixed at 0.3 when $\beta_{GE2}$ was positive and -0.3 when $\beta_{GE2}$ was negative.

For model (iii), $Y$ is a function of only the interaction between $G$ and $E_1$. For this model, the interaction effect $\beta_{GE1}$ and exposure frequency were chosen such that the observed marginal effect of $G$ was fixed at 10% of the trait standard deviation.

The corresponding working association (testing) model was $E[Y] = \beta_G G$ (analogous to equation (2) in Material and Methods) in all cases since information on $E_1$ and $E_2$ was assumed to be unavailable.

To assess the type 1 error level of the joint location-scale methods at 0.05, 0.005 and 0.0005 levels, we simulated 100,000 replicate samples of $n = 2000$ subjects each from the null genetic model with no genetic association (i.e. $\beta_G = 0$ and $\beta_{GE} = 0$). (Results of $n = 1000$ and 4000 are characteristically similar.) To examine the behavior of the testing methods under small group sizes, we conducted additional simulations under varied MAF (0.3, 0.2, 0.1, 0.05 and 0.03) as well as under fixed genotype group sizes where the rare homozygote group size was small (2, 5, 7, 10, 15 or 20) with the other genotype group sizes determined with respect to Hardy Weinberg equilibrium. For comparison, empirical type 1 error rates of the individual location-only and scale-only tests are also included in Tables 1 and 2, in addition to the JLS-Fisher and JLS-minP tests, and the LRT of Cao, et al. [2014]; the distribution test of Aschard, et al. [2013] has the correct type 1 error by design. Type 1 error control at the genome-wide level was assessed by phenotype-permutation analysis of two application data (results not shown).

For power evaluation, as in Aschard, et al. [2013], we focused on MAF = 0.3 and $n = 2000$ for models (i) and (ii) and n=4000 for model (iii). Power (at the $5 \times 10^{-8}$ level) was estimated from 500 replicates, based on asymptotic p-values of the tests considered (with the exception of the distribution test). For the distribution test, p-values required estimation by simulations and corresponding power results were from Aschard, et al. [2013], kindly provided by Drs. Aschard and Kraft.

*Results*

Investigation of type 1 error (100,000 replicates) under the normal model showed that all individual and joint tests considered here maintained the correct error rates at the 0.05, 0.005, and 0.0005 levels when MAF was at least 0.1 and there were at least 20 individuals within each genotype group (Tables 1 and 2). However, as MAF or the smallest genotype group size decreased (<20), the LRT method demonstrated inflated type 1 error (Tables 1 and 2). Departure from normality also resulted in LRT having inflated type 1 error, consistent and as previously discussed in Cao, et al. [2014].

Under the normal model, the LRT and JLS-Fisher methods had similar power and were more powerful than the JLS-minP and distribution tests (Figures 1 and 2), and the location-only and scale-only tests (Figures 3 and 4), in most scenarios considered. For example, for Model (i) when there were both main and interaction effects, and $\beta_G$ = 0.01, $\beta_{E1}$ = 0.3 and $\beta_{GE1}$ = 0.6 (and MAF = 0.3, prob($E_1$=1)=0.3 and $n$ = 2000), power of the location-only, scale-only, distribution, LRT, JLS-minP, and JLS-Fisher tests were, respectively, 0.25, 0.05, 0.36, 0.69, 0.24 and 0.66 (Figures 1a and 3a; $\alpha$ = 5 · $10^{-8}$). For the same model but when the interaction effect is in opposite direction to the main genetic effect, $\beta_G$ = 0.05, $\beta_{E1}$ = −0.3, $\beta_{GE1}$ = −0.6, power were respectively 0.01, 0.03, 0.05, 0.24, 0.03 and 0.22 (Figures 1e and 3e).

For Model (ii) where the main $G$ effect was excluded but two main $E$ effects and two $G$x$E$ interaction effects were included, the findings were similar (Figures 2 and 4); JLS-Fisher and LRT were most powerful in the presence of multiple interaction effects in either the same, or opposite direction.

In Model (iii) with consideration of a more extreme scenario, when there were no main $G$ or $E$ effects while the interaction effect was large ($\beta_{GE1}$ = 2) and the unobserved exposure was rare (prob($E_1$ = 1) = 0.05), the distribution test was more powerful than the JLS-Fisher test (0.916 vs. 0.406) (row 1 of Table 3). This is because the resulting phenotype distributions across genotypes differed in shape and their differences were not well captured by only mean and variance parameters. (The JLS-Fisher test could, in theory, be extended to include the skewness parameter or higher moments of such a distribution. The gain in power for this particular setting, however, would be at the expense of power loss for other approximately normally distributed traits.) In this scenario, when the phenotype deviated from normality, the LRT method appeared to be most powerful (power = 0.996). However, further analysis using permutation estimation of p-values showed that power of the LRT under asymptotic analysis was greatly inflated, while the proposed JLS methods were robust (Table 4).

**Discussion**

The grouping of genes into gene sets and pathways, implicitly assumes interactions between the group members. Yet standard gene-set and pathway analysis tools in large-part do not account for these relationships. Here we provide a method to account for these relationships, while avoiding specification of the contributing interactions.

The alternative distribution [Aschard, et al. 2013] and LRT [Cao, et al. 2014] joint testing methods were proposed for analysis of single SNPs. In principle, they can be extended to gene-set analysis, however, multiple issues arise in implementation. The distribution test statistic depends on the size of the genotype groups, so it is not clear what the best strategy is to combine the statistics across SNPs with different MAFs. The LRT method is sensitive to the normality assumption of the phenotype distribution and to small group size of the genotype distribution (Tables 1 and 2).

JLS testing is also extremely relevant for single variant association studies. In the GWAS or next-generation sequencing (NGS) settings where millions or tens of millions of SNPs are investigated, rapid screening of the whole genome to correctly prioritize SNPs for further examination demands methods that are powerful yet computationally efficient. The proposed JLS testing method is robust and easy to implement, suitable for large-scale whole-genome scans, and can reveal individual genetic variants with main and/or interaction effects without the need to explicitly specify the interacting genetic and/or environmental variables. Compared with the distribution test and LRT alternatives, our method combines both simplicity of implementation and robustness to small size of the rare homozygous genotype group. A step-by-step guide for application of the JLS framework presented in this paper is provided in the Appendix.

Violations of the normal data assumption can affect the type 1 error level of the proposed JLS test method as it does the LRT approach (increased type 1 error rate is more severe with LRT). This is largely due to the assumption of independence between the individual location-only and scale-only tests, which is required for the $\chi^2_4$ approximation of the JLS-Fisher statistic. To circumvent this issue, investigators may choose to rely on a permutation distribution when estimating p-values. However, this aspect may impose computation challenges at the genome-wide level. An alternative approach is to model the dependency between the individual location and scale test statistics and obtain an adjusted distribution for the JLS statistic. These adjusted distributions have been explored elsewhere under complete specification of the dependency structure between the test statistics being combined [Chuang and Shih 2012; Hou 2005; Kost and McDermott 2002]. However, it remains to be explored how to model the dependency between individual location and scale association test statistics at different loci across the genome.

Recent literature cautions the investigation and interpretation of significant differences in phenotypic variability across genotypes as well as observed GxE and GxG interaction effects [Cao, et al. 2014; Dudbridge and Fletcher 2014; Struchalin, et al. 2010; Sun, et al. 2013; Wood, et al. 2014]. Firstly, variance heterogeneity across genotypes can result from a mean-variance relationship induced by an inappropriate choice of scale for the phenotype under investigation. For example, if the true underlying model is strictly due to a main effect, $Y \sim \alpha + \beta_G G + \varepsilon$, but $\varepsilon \sim$ logNormal$(0,\sigma^2)$, the researcher may analyze a logarithmic transformation of *Y* for its statistical properties. Such scaling can also inadvertently occur when the choice of phenotype for statistical analysis does not directly represent the true underlying biological outcome of interest (which we may not be able to recognize or observe directly). Consequently, this could result in observed phenotypic mean and variance differences between genotypes, where the variance difference is an artifact not due to a genetic interaction. Secondly, dependence between a causal loci and an exposure ($G$ or $E$) can create synthetic observed interactions between the exposure and other tag SNPs in incomplete LD with the causal loci [Dudbridge and Fletcher 2014; Wood, et al. 2014]. Consequently, we may observe variance heterogeneity of the phenotype across tag SNP genotypes [Cao, et al. 2014]. In these scenarios, however, since there is a true underlying genotype-phenotype relationship (main effect of $G$), it does not matter if we detect this association through a location or a scale test, or both; the goal of our approach is to identify the associated $G$ and generate hypotheses, but not to determine the precise biological mechanism. In light of this, the joint approach to variant detection bypasses the difficulty of ensuring the proper scale of measurement on the phenotype and reduces the concerns for decreased power when analyzing tag SNPs as opposed to actual causal SNPs or for declaring spurious/synthetic interaction as causal interaction.

In consideration of the choices available for the individual location and scale tests and the methods of combining information from these individual components, we recognize that there is no single, most powerful method for all circumstances [Derkach, et al. 2013; Owen 2009]. We show that, for normally distributed traits, the JLS-Fisher test statistic $W_F$ follows a $\chi^2_4$ distribution under the null hypothesis, as long as the location test statistic is a function of the complete sufficient statistic (e.g. linear regression t-statistic, ANOVA F-statistic) and the distribution of the scale test statistic does not depend on the model parameters (e.g. Levene's test or the F-test for equality of variances). In practice, when the normality assumption is violated or other JLS tests are preferred, permutation-based p-value evaluation can be used with increased computational cost.

Challenges in replicating original findings, particularly in the GxG, GxE, gene-set and pathway analysis setting, is well documented [Hutter, et al. 2013]. Given the potential heterogeneity between the Canadian discovery sample and the French replication sample (e.g. clinical treatment, climate and air quality), interactions could differ between the two samples. Therefore, even if the exposure variables were precisely measured, a significant multivariate interaction effect found in the discovery sample may not be observed in the replication sample. The proposed method allows different exposure variables between genes and between samples, as long as the corresponding underlying phenotype-genotype association mechanism results in phenotypic mean and/or variance differences between genotypes of the variants of interest.

The proposed framework can be easily extended for meta-analysis, where the sample- or study-specific association test statistics or p-values to be combined across samples are obtained from JLS testing applications instead of the typical location-only testing method. In this case, the sample-specific choices of the individual location and scale tests need not be identical across different studies, as long as p-values of the JLS tests are valid within each study. Analyzing imputed SNPs or correlated samples (i.e. pedigree data) using the proposed JLS framework would require development of appropriate scale-only test methods; location-only test methods for these more complex settings are already available [Acar and Sun 2013; Aulchenko, et al. 2010; Horvath, et al. 2001; Lou, et al. 2008; Schaid, et al. 2002; Yu 2012].

In conclusion, we have provided a robust joint location-scale testing framework for the detection of single variant, gene-set or pathway associations involving either main or interaction effects, or both, with a quantitative trait, regardless of the biological interpretation of the chosen scale of phenotypic measurement. This method should be considered for future gene-set, pathway and whole-genome association scans and be employed to re-examine datasets previously analyzed using the conventional location-only or scale-only testing approaches, for complex traits or complex secondary phenotypes of 'simple' Mendelian diseases. The method will help researchers pinpoint susceptibility loci for additional analysis towards the understanding of the genetic architecture of a complex trait.

**Acknowledgements**

This work was funded by the Natural Sciences and Engineering Research Council of Canada (NSERC; 250053-2013 to L.S.); the Canadian Institutes of Health Research (CIHR; 201309MOP-310732-G-CEAA-117978 to L.S. and MOP-258916 to L.J.S.); A.D.P. held a Canada Research Chair in the Genetics of Complex Diseases. D.S. is a trainee of CIHR STAGE (Strategic Training for Advanced Genetic Epidemiology) program at the University of Toronto. The authors also thank Drs. Hughes Aschard and Peter Kraft for providing their simulation results.

**Appendix**
*Implementation Steps for Applying the JLS Testing Framework*

Note that the following implementation is appropriate for testing association of a phenotype with a genotyped genetic variant (e.g. SNP) using a sample of unrelated subjects.

1. Check the phenotype of interest for fit to a normal distribution. If required, adjust the phenotype using a suitable transformation, e.g. inverse normal transform. If the researcher proceeds using a non-normal phenotype, only the permutation (resampling-based) p-value analysis will be valid (see step 4 (b)).

2. Choose the individual location and scale tests based on the distribution of phenotype (normal or non-normal) or preference, for example parametric or non-parametric versions of each test. In the present paper, our phenotypes were normally distributed (after transformation) and we chose linear regression and Levene's test for the location and scale tests, respectively.

3. Choose a JLS testing method of combining information from the individual location and scale tests and calculate the JLS test statistic. We acknowledge that there is no 'most powerful' method for all situations in practice. Based on our experience, we recommend the use of Fisher's method (JLS-Fisher) of combining the association evidence:

$$W_F = -2\log(p_L) - 2\log(p_S),$$

where $p_L$ and $p_S$ are the individual location and scale test p-values, respectively.

4. Chose the p-value estimation method for the JLS statistic:

   (a) Based on the approximate asymptotic distribution of the JLS test statistic. For the JLS-Fisher example, $W_F$ is distributed as a $\chi_4^2$ random variable, if the chosen individual location and scale tests are independent of each other under the null hypothesis; this assumption is correct if the trait is normally distributed and if the location-only test statistic is a function of the complete sufficient statistic (e.g. linear regression t-statistic, ANOVA F-statistic) and the distribution of the scale-only test statistic does not depend on the model parameters (e.g. Levene's test or the F-test for equality of variances).

(b) Based on resampling methods such as permutation:
- Calculate the observed JLS test statistic, e.g. $W_F$.
- Choose the number of permutation replicates, $K$, based on the desired p-value accuracy.
- Permute the phenotype independently $K$ times, and for each replicate $k$, recalculate the JLS test statistic, $W_F^k$, $k$ = 1, …, $K$.
- Obtain the permutation p-value as [the number of $W_F^K > W_F$]/K.

**Figures and Tables**

*Figures*

- Figure 1. Power comparison under simulation model [i].
- Figure 2. Power comparison under simulation model [ii] – Proposed and competing joint location-scale testing methods.
- Figure 3. Power comparison under simulation model [i] – Proposed joint location-scale testing method and individual location-only or scale-only testing methods.
- Figure 4. Power comparison under simulation model [ii] – Proposed joint location-scale testing method and individual location-only or scale-only testing methods.

*Tables*

- Table 1. Type 1 error comparison with varied minor allele frequency (MAF).
- Table 2. Type 1 error comparison with varied genotypic group sizes.
- Table 3. Power comparison under simulation model [iii].
- Table 4. Power comparison under simulation model [iii] – p-value estimation based on approximate distribution of the test statistics vs. permutation-based p-value estimation.

**Figure 1. Power comparison under simulation model [i].** Four different testing methods are examined: the proposed JLS-Fisher (red) and JLS-minP (purple) methods, the distribution test (blue) of Ashcard, et al. (2013), and the LRT (black) of Cao et al. (2014). Phenotype values for 2000 independent subjects were simulated under $E[Y] = \beta_G G + \beta_{E1} E_1 + \beta_{GE1} G \cdot E_1$, where the MAF of $G$ was 0.3 and the exposure variable $E_1$ was simulated as a Bernoulli variable with frequency=0.3. The effect of the exposure $\beta_{E1}$ was fixed at 0.3 while the other effects vary. Top panel (A)-(C) are results when the main genetic effect $\beta_G$ and the interaction effect $\beta_{GE1}$ are in the same direction, and the bottom panel (E)-(F) are results when $\beta_G$ and $\beta_{GE1}$ are in opposite direction. Power was calculated at the $5 \times 10^{-8}$ level based on 500 replicates.

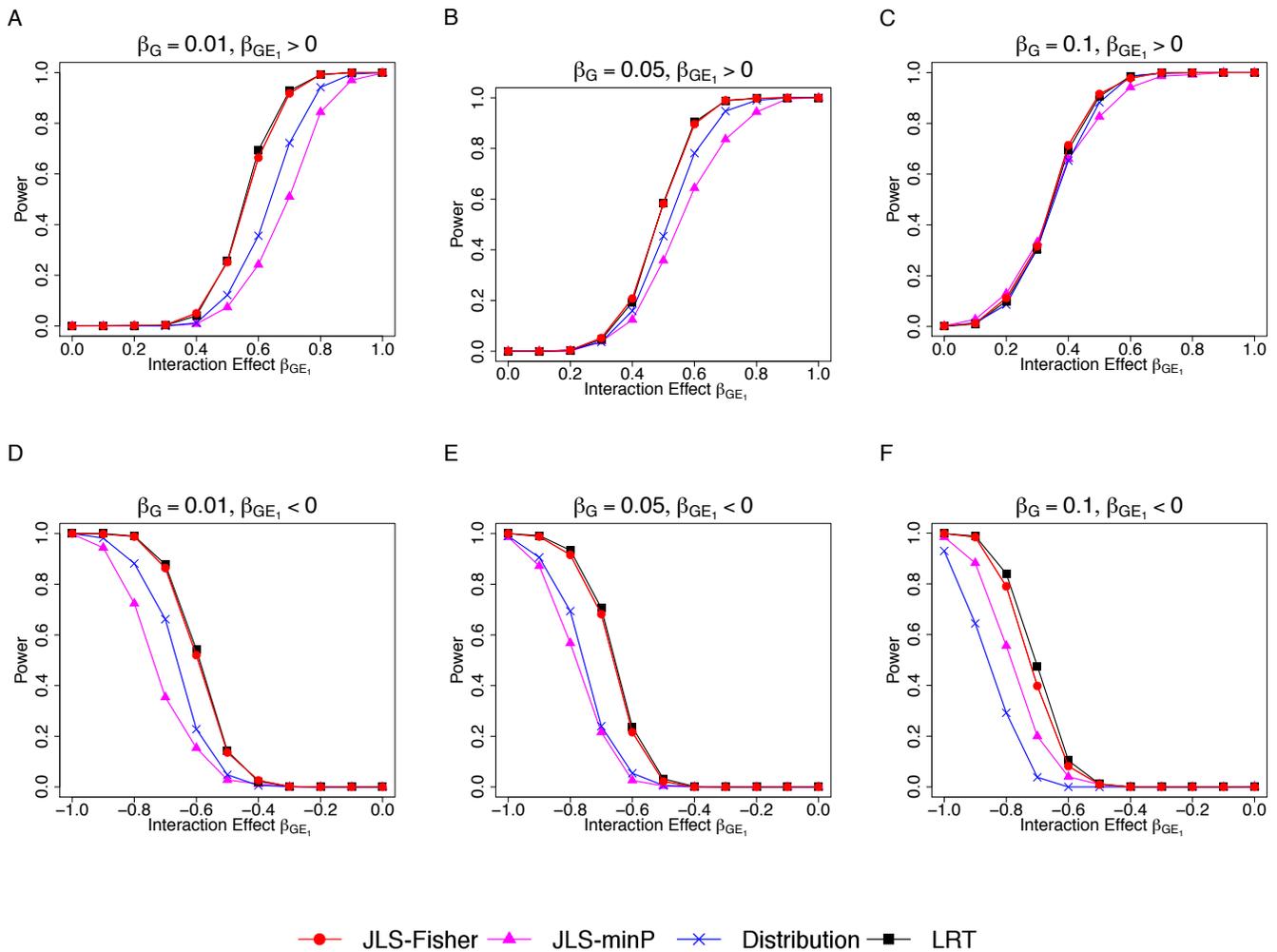

**Figure 2. Power comparison under simulation model [ii] – Proposed and competing joint location-scale testing methods.** Four different joint location-scale testing methods are examined: the proposed JLS-Fisher (red) and JLS-minP (purple) tests, and the distribution test (blue) of Ashcard, et al. [2013] and the LRT (black) of Cao, et al. [2014]. Phenotype values for 2000 independent subjects were simulated under $E[Y] = \beta_{E1}E_1 + \beta_{E2}E_2 + \beta_{GE1}G \cdot E_1 + \beta_{GE2}G \cdot E_2$, where the MAF of $G$ was 0.3 and the exposure variables $E_1$ and $E_2$ were simulated as Bernoulli variables with frequency=0.3. The effect of the exposure $\beta_{E1}$ was fixed at 0.3 while the interaction effect $\beta_{GE1}$ varied. The effect of exposure $\beta_{E2}$ was fixed at 0.3 when the interaction effect $\beta_{GE2}$ was positive, and -0.3 when $\beta_{GE2}$ was negative. Results are presented for models when effects of the two interaction effects, $\beta_{GE1}$ and $\beta_{GE2}$, are in the same direction (A), and when the two interaction effects are in opposite direction having different amplitude (B) or the same amplitude (C). Power was calculated at the $5 \times 10^{-8}$ level based on 500 replicates.

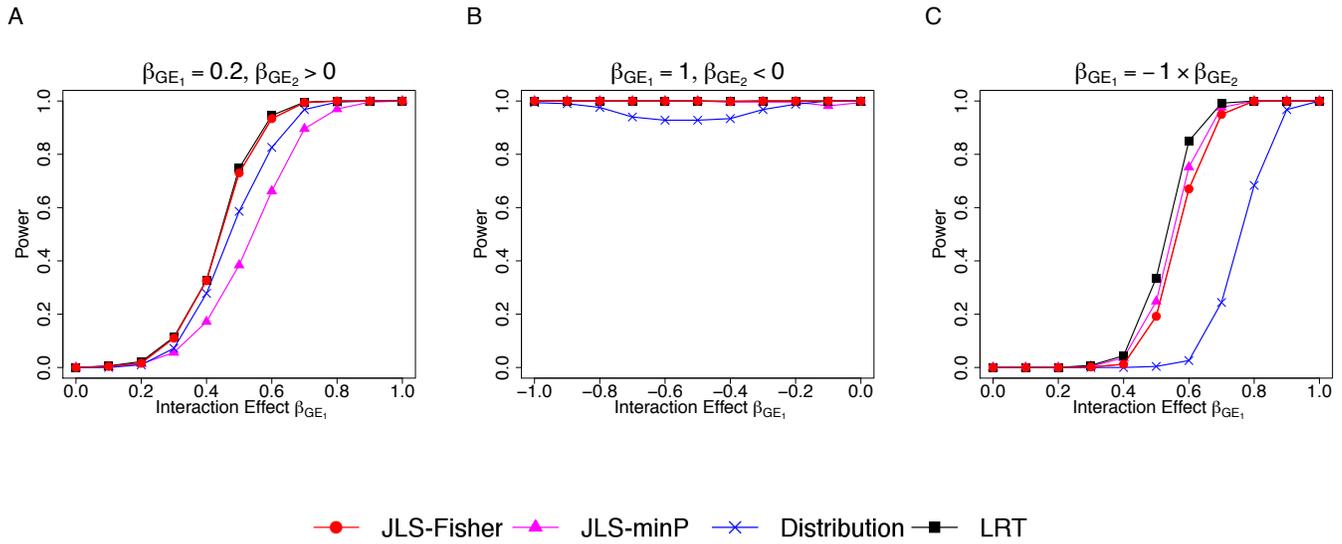

**Figure 3. Power comparison under simulation model [i] – Proposed joint location-scale testing method and individual location-only or scale-only testing methods.** The proposed JLS-Fisher test (red) is compared to the individual regression location-only test (orange) and Levene's scale-only test (green). Phenotype values for 2000 independent subjects were simulated under $E[Y] = \beta_G G + \beta_{E1} E_1 + \beta_{GE1} G \cdot E_1$, where the MAF of $G$ was 0.3 and the exposure variable $E_1$ was simulated as a Bernoulli variable with frequency=0.3. The effect of the exposure $\beta_{E1}$ was fixed at 0.3 while the other effects vary. Top panel (A)-(C) are results when the main genetic effect $\beta_G$ and the interaction effect $\beta_{GE1}$ are in the same direction, and the bottom panel (D)-(F) are results when $\beta_G$ and $\beta_{GE1}$ are in opposite direction. Power was calculated at the $5 \times 10^{-8}$ level based on 500 replicates.

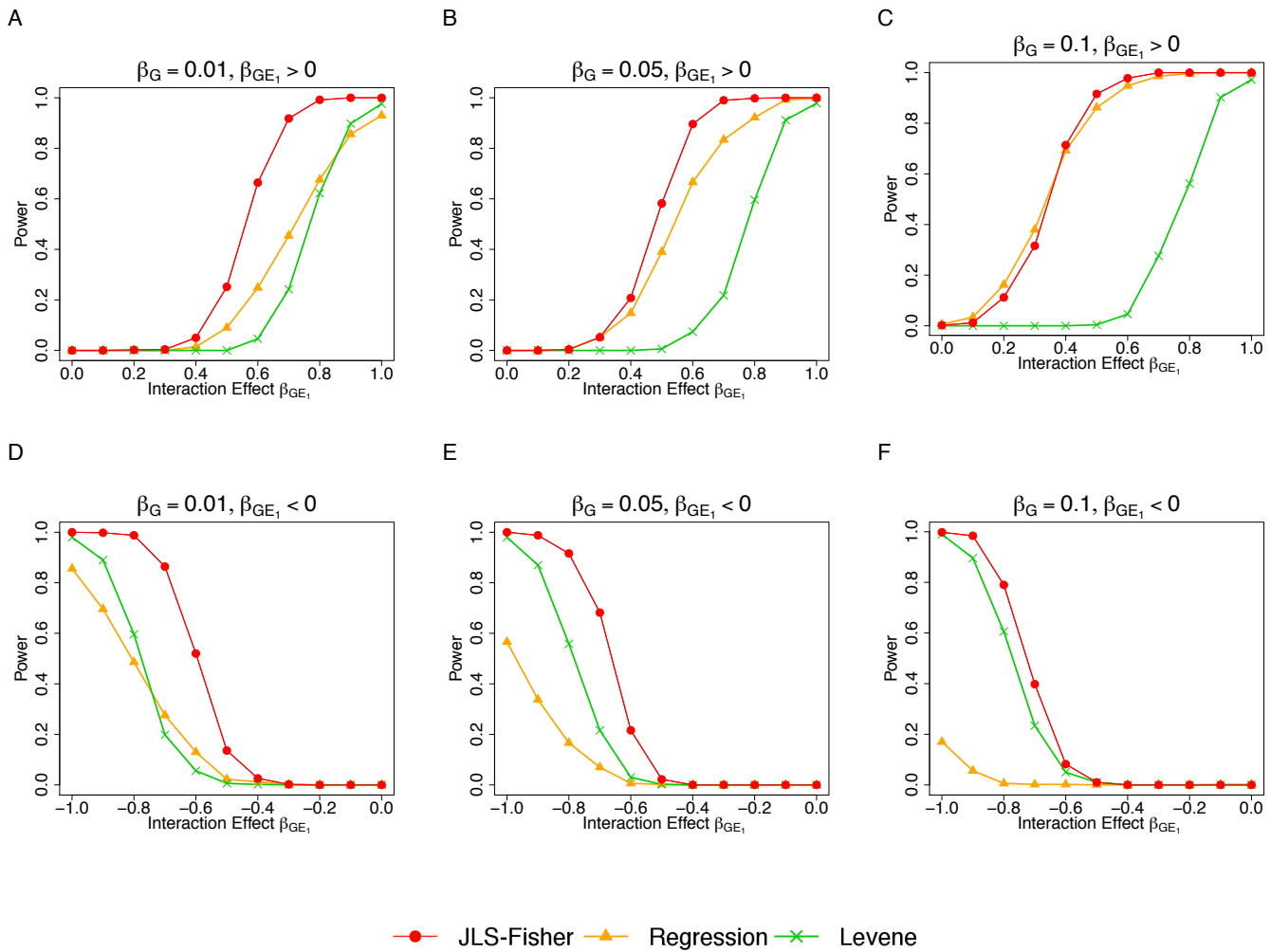

**Figure 4. Power comparison under simulation model [ii] – Proposed joint location-scale testing method and individual location-only or scale-only testing methods.** The proposed JLS-Fisher test (red) is compared to the individual regression location-only test (orange) and Levene's scale-only test (green). Phenotype values for 2000 independent subjects were simulated under $E[Y] = \beta_{E1}E_1 + \beta_{E2}E_2 + \beta_{GE1}G \cdot E_1 + \beta_{GE2}G \cdot E_2$, where the MAF of $G$ was 0.3 and the exposure variables $E_1$ and $E_2$ were simulated as Bernoulli variables with frequency=0.3. The effect of the exposure $\beta_{E1}$ was fixed at 0.3 while the interaction effect $\beta_{GE1}$ varied. The effect of exposure $\beta_{E2}$ was fixed at 0.3 when the interaction effect $\beta_{GE2}$ was positive, and -0.3 when $\beta_{GE2}$ was negative. Results are presented for models when effects of the two interaction effects, $\beta_{GE1}$ and $\beta_{GE2}$, are in the same direction (A), and when the two interaction effects are in opposite direction having different amplitude (B) or the same amplitude (C). Power was calculated at the $5 \times 10^{-8}$ level based on 500 replicates.

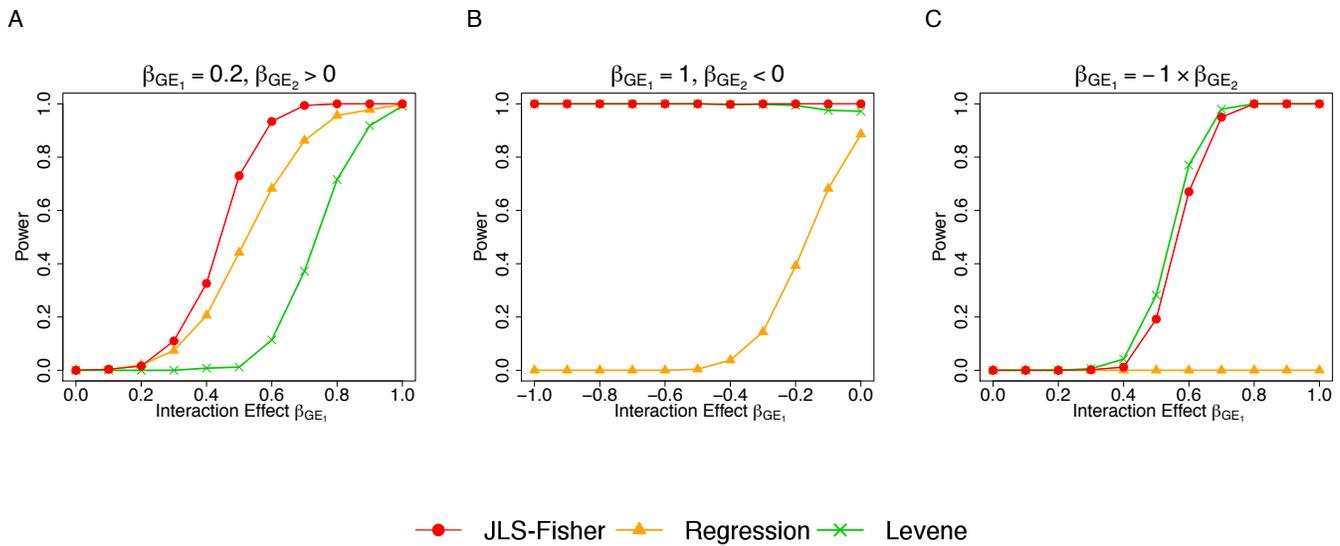

**Table 1. Type 1 error comparison with varied minor allele frequency (MAF).**
Type 1 error is presented for the regression location test (Reg), Levene's scale test (Levene), the proposed JLS-Fisher and JLS-minP tests, and the LRT of Cao, et al. [2014]. The distribution test of Ashcard, et al. [2013] has correct type 1 error by design. Phenotype values for 2,000 independent subjects were simulated under the null genetic model with no genetic association (model [i] with $\beta_G = 0, \beta_E = 0, \beta_{GE} = 0$) and varied MAF (0.03-0.3) with residual variation from a normal distribution with mean 0 and standard deviation 1. Statistical significance for each replicate was calculated at the 0.05, 0.005, and 0.0005 levels based on 100,000 replicates.

| MAF=0.3 | | | | | |
|---|---|---|---|---|---|
| | Individual | | Joint | | |
| level | Reg | Levene | JLS-Fisher | JLS-MinP | LRT |
| 0.05 | 0.05005 | 0.04988 | 0.04931 | 0.04952 | 0.05144 |
| 0.005 | 0.00490 | 0.00497 | 0.00476 | 0.00488 | 0.00473 |
| 0.0005 | 0.00047 | 0.00056 | 0.00048 | 0.00045 | 0.00045 |

| MAF=0.2 | | | | | |
|---|---|---|---|---|---|
| | Individual | | Joint | | |
| level | Reg | Levene | JLS-Fisher | JLS-MinP | LRT |
| 0.05 | 0.04983 | 0.04788 | 0.04869 | 0.04910 | 0.05060 |
| 0.005 | 0.00475 | 0.00473 | 0.00508 | 0.00481 | 0.00476 |
| 0.0005 | 0.00049 | 0.00054 | 0.00053 | 0.00058 | 0.00047 |

| MAF=0.1 | | | | | |
|---|---|---|---|---|---|
| | Individual | | Joint | | |
| level | Reg | Levene | JLS-Fisher | JLS-MinP | LRT |
| 0.05 | 0.05101 | 0.04766 | 0.04888 | 0.04860 | 0.05254 |
| 0.005 | 0.00476 | 0.00435 | 0.00460 | 0.00452 | 0.00506 |
| 0.0005 | 0.00047 | 0.00058 | 0.00049 | 0.00049 | 0.00053 |

| MAF=0.05 | | | | | |
|---|---|---|---|---|---|
| | Individual | | Joint | | |
| level | Reg | Levene | JLS-Fisher | JLS-MinP | LRT |
| 0.05 | 0.05026 | 0.04183 | 0.04603 | 0.04624 | 0.06130 |
| 0.005 | 0.00473 | 0.00498 | 0.00460 | 0.00524 | 0.00777 |
| 0.0005 | 0.00047 | 0.00064 | 0.00060 | 0.00063 | 0.00107 |

| MAF=0.03 | | | | | |
|---|---|---|---|---|---|
| | Individual | | Joint | | |
| level | Reg | Levene | JLS-Fisher | JLS-MinP | LRT |
| 0.05 | 0.05026 | 0.04788 | 0.04900 | 0.04959 | 0.06722 |
| 0.005 | 0.00523 | 0.00583 | 0.00568 | 0.00592 | 0.01120 |
| 0.0005 | 0.00045 | 0.00110 | 0.00089 | 0.00087 | 0.00231 |

**Table 2. Type 1 error comparison with varied genotypic group sizes.** Type 1 error is presented for the regression location test (Reg), Levene's scale test (Levene), the proposed JLS-Fisher and JLS-minP tests, and the LRT of Cao, et al. [2014]. The distribution test of Ashcard, et al. [2013] has correct type 1 error by design. Phenotype values for 2,000 independent subjects were simulated under the null genetic model with no genetic association (model [i] with $\beta_G = 0, \beta_E = 0, \beta_{GE} = 0$) with residual variation from a normal distribution with mean 0 and standard deviation 1. The genotype group sizes were fixed with respect to the smallest group size ($N_{smallest}$=2, 5, 7, 10, 15 or 20) and the corresponding Hardy Weinberg equilibrium. Statistical significance for each replicate was calculated at the 0.05, 0.005, and 0.0005 levels based on 100,000 simulation replicates.

| Group Sizes ($N_0, N_1, N_2$)=(1882,116,2) (MAF~0.03) | | | | | |
|---|---|---|---|---|---|
| | Individual | | Joint | | |
| level | Reg | Levene | JLS-Fisher | JLS-MinP | LRT |
| 0.05 | 0.04890 | 0.05369 | 0.05350 | 0.05593 | 0.09134 |
| 0.005 | 0.00485 | 0.01002 | 0.00820 | 0.00862 | 0.02108 |
| 0.0005 | 0.00045 | 0.00221 | 0.00157 | 0.00166 | 0.00582 |

| Group Sizes ($N_0, N_1, N_2$)=(1805,190,5) (MAF~0.05) | | | | | |
|---|---|---|---|---|---|
| | Individual | | Joint | | |
| level | Reg | Levene | JLS-Fisher | JLS-MinP | LRT |
| 0.05 | 0.05039 | 0.03769 | 0.04263 | 0.04262 | 0.05869 |
| 0.005 | 0.00496 | 0.00356 | 0.00399 | 0.00411 | 0.00712 |
| 0.0005 | 0.00043 | 0.00043 | 0.00036 | 0.00042 | 0.00079 |

| Group Sizes ($N_0, N_1, N_2$)=(1767,226,7) (MAF~0.06) | | | | | |
|---|---|---|---|---|---|
| | Individual | | Joint | | |
| level | Reg | Levene | JLS-Fisher | JLS-MinP | LRT |
| 0.05 | 0.05087 | 0.04103 | 0.04578 | 0.04591 | 0.05677 |
| 0.005 | 0.00505 | 0.00406 | 0.00451 | 0.00488 | 0.00617 |
| 0.0005 | 0.00041 | 0.00061 | 0.00047 | 0.00063 | 0.00070 |

| Group Sizes ($N_0, N_1, N_2$)=(1730,320,10) (MAF~0.07) | | | | | |
|---|---|---|---|---|---|
| | Individual | | Joint | | |
| level | Reg | Levene | JLS-Fisher | JLS-MinP | LRT |
| 0.05 | 0.04999 | 0.04524 | 0.04776 | 0.04709 | 0.0542 |
| 0.005 | 0.00474 | 0.00506 | 0.00487 | 0.00502 | 0.00571 |
| 0.0005 | 0.00037 | 0.00071 | 0.00067 | 0.00065 | 0.00071 |

| Group Sizes ($N_0, N_1, N_2$)=(1674,311,15) (MAF~0.085) | | | | | |
|---|---|---|---|---|---|
| | Individual | | Joint | | |
| level | Reg | Levene | JLS-Fisher | JLS-MinP | LRT |
| 0.05 | 0.05037 | 0.04539 | 0.04808 | 0.04845 | 0.05263 |
| 0.005 | 0.00486 | 0.00509 | 0.00471 | 0.00464 | 0.00553 |
| 0.0005 | 0.00046 | 0.00055 | 0.00050 | 0.00051 | 0.00061 |

| Group Sizes ($N_0, N_1, N_2$)=(1620,360,20) (MAF~0.1) | | | | | |
|---|---|---|---|---|---|
| | Individual | | Joint | | |
| level | Reg | Levene | JLS-Fisher | JLS-MinP | LRT |
| 0.05 | 0.05090 | 0.04704 | 0.04953 | 0.04936 | 0.05319 |
| 0.005 | 0.00514 | 0.00414 | 0.00470 | 0.00472 | 0.00500 |
| 0.0005 | 0.00050 | 0.00057 | 0.00046 | 0.00060 | 0.00045 |

**Table 3. Power comparison under simulation model [iii].** Power is presented for the regression location test (Reg), Levene's scale test (Levene), the proposed JLS-Fisher and JLS-minP tests, the distribution (Dist.) test of Ashcard, et al. [2013], and the LRT of Cao, et al. [2014]. Phenotype values for 4000 independent subjects were simulated under E[$Y$] = $\beta_{GE1}G \cdot E_1$, where the MAF of $G$ was 0.3. Results are presented for models when the interaction effect, $\beta_{GE1}$, and exposure ($E_1$) frequency were chosen such that the observed marginal effect of G was fixed at 10% of the trait standard deviation. Power was calculated at the $5\times10^{-8}$ level based on 500 replicates.

|  |  | Individual | | Joint | | | |
|---|---|---|---|---|---|---|---|
| Freq-$E_1$ | Int. Effect ($\beta_{GE1}$) | Reg. | Levene | JLS-Fisher | JSL-minP | LRT | Dist. |
| 0.05 | 2 | 0.010 | 0.110 | 0.406 | 0.090 | 0.996 | 0.916 |
| 0.1 | 1 | 0.040 | 0.040 | 0.378 | 0.054 | 0.740 | 0.178 |
| 0.2 | 0.5 | 0.038 | 0.000 | 0.098 | 0.026 | 0.110 | 0.064 |
| 0.3 | 0.33 | 0.110 | 0.000 | 0.086 | 0.084 | 0.084 | 0.028 |
| 0.5 | 0.2 | 0.080 | 0.000 | 0.046 | 0.064 | 0.036 | 0.054 |
| 1 | 0.1 | 0.076 | 0.000 | 0.042 | 0.060 | 0.036 | 0.040 |

**Table 4. Power comparison under simulation model [iii] – p-value estimation based on approximate distribution of the test statistics vs. permutation-based p-value estimation.** Power is presented for the regression location-only test (Reg), Levene's scale-only test (Levene), the proposed JLS-Fisher and JLS-minP tests, the distribution (Dist.) test of Ashcard, et al. [2013], and the LRT of Cao, et al. [2014]. Phenotype values for 1000 independent subjects were simulated under E[$Y$] = $\beta_{GE1}G \cdot E_1$, where the MAF of $G$ was 0.3 and the exposure variable $E_1$ was simulated as a Bernoulli variable with frequency=0.05. The interaction effect $\beta_{GE1}$ was fixed at 2 while the other effects varied. Power was calculated at the 0.01 significance level based on 500 replicates. For each replicate, permutation p-values were estimated from 10,000 iterations.

|  | Individual | | Joint | | | |
|---|---|---|---|---|---|---|
| P-value estimation method | Reg. | Levene | JLS-Fisher | JLS-minP | LRT | Dist. |
| Asymptotic | 0.190 | 0.266 | 0.390 | 0.288 | 0.932 | NA |
| Permutation | 0.192 | 0.268 | 0.364 | 0.288 | 0.698 | 0.812 |